# ELM control with RMP: plasma response models and the role of edge peeling response

Yueqiang Liu[1,2,3,*], C.J. Ham[1], A. Kirk[1], Li Li[4,5,6], A. Loarte[7], D.A. Ryan[8,1], Youwen Sun[9], W. Suttrop[10], Xu Yang[11], Lina Zhou[11]

[1]CCFE, Culham Science Centre, Abingdon, OX14 3DB, UK
[2]Southwestern Institute of Physics, PO Box 432, Chengdu 610041, People's Republic of China
[3]Department of Earth and Space Science, Chalmers University of Technology, SE-412 96 Gothenburg, Sweden
[4]College of Science, Donghua University, Shanghai 201620, China
[5]Forschungszentrum Jülich GmbH, Institut für Energie- und Klimaforschung - Plasmaphysik, Jülich, Germany
[6]Member of Magnetic Confinement Fusion Research Centre, Ministry of Education, China
[7]ITER Organization, Route de Vinon-sur-Verdon, CS 90 046, 13067 St Paul-lez-Durance Cedex, France
[8]Department of Physics, University of York, Heslington, York, YO10 5DD, UK
[9]Institute of Plasma Physics, Chinese Academy of Sciences, PO Box 1126, Hefei 230031, People's Republic of China
[10]Max-Planck-Institut for Plasmaphysik, D-85740 Garching, Germany
[11]Key Laboratory of Materials Modification by Laser, Ion and Electron Beams, Ministry of Education, School of Physics and Optoelectronic Technology, Dalian University of Technology, Dalian 116024, People's Republic of China

[*]E-mail: yueqiang.liu@ukaea.uk

## Abstract

Resonant magnetic perturbations (RMP) have extensively been demonstrated as a plausible technique for mitigating or suppressing large edge localized modes (ELMs). Associated with this is a substantial amount of theory and modelling efforts during recent years. Various models describing the plasma response to the RMP fields have been proposed in the literature, and are briefly reviewed in this work. Despite their simplicity, linear response models can provide alternative criteria, than the vacuum field based criteria, for guiding the choice of the coil configurations to achieve the best control of ELMs. The role of the edge peeling response to the RMP fields is illustrated as a key indicator for the ELM mitigation in low collisionality plasmas, in various tokamak devices.

## 1. Introduction

Among many macroscopic MHD instabilities, the type-I edge localized mode (ELM) is one of the most important and special ones in tokamak fusion plasmas. Unlike the neoclassical tearing mode or the resistive wall mode, that poses pressure limits on the plasmas, and can eventually lead to major disruptions of the discharges, type-I ELMs, which represent a periodic release of the plasma energy, do not generally limit the plasma pressure. Unlike the internal kink mode whose influence is typically confined to the plasma core region, ELMs are born in the plasma edge region – the so called pedestal region of the H-mode plasma – and as such, have direct consequences on the plasma-wall interactions. Whilst frequent ELMs may be beneficial in flushing out impurities from the plasma, large, less frequent ELM bursts can generate a substantial amount of heat flux reaching the plasma facing components in tokamaks, potentially causing significant damage to the materials. In fact this damage is shown to be at an unacceptable level for next generation fusion devices such as ITER [1].

Among several ELM control techniques, the resonant magnetic perturbation (RMP), produced by magnetic coils outside the plasma, appears to be a robust one. Several tokamak devices have experimented with this technique to control type-I ELMs, including DIII-D [2], JET [3], MAST [4], ASDEX Upgrade [5], KSTAR [6], and recently EAST [7]. Either ELM mitigation (increasing the ELM frequency accompanied by reduction of the individual ELM amplitude) or ELM suppression (complete removal of type-I ELMs) has been achieved on these devices, using the RMP fields.

Figure 1 shows a schematic view of RMP coils used for ELM control. Typically there are several rows (between 1-3 in most of the devices) of coils along the poloidal angle, with each row consisting of a number of coils distributed along the toroidal angle of the torus. Two rows of coils are shown in the figure, as an example. Figure 1(b) shows a plane-view of the coils geometry, along the toroidal (horizontal axis) and poloidal (vertical axis) angles. Typically the periodic distribution of the window-frame coils along the toroidal angle allows generation of the vacuum magnetic field of different (dominant) toroidal mode numbers $n$. For the case shown in Fig. 1, with 8 coils uniformly distributed along the toroidal angle, it is possible to generate the $n$=1,2, or 4 components with small sidebands. It is of course also possible to generate the *n*=3 fields with 8 coils, but with a large sideband of *n*=5. The other parameters that determine the coils geometry, as well as the toroidal phase of the coil currents, are the toroidal location $\phi_j$ of the centre of each coil, the toroidal coverage of each coil $\Delta\phi_j$ which is normally the same for all coils, the separation angle $\delta\phi$ between two neighbouring coils, and a reference angle $\phi_0$ for the coil. These parameters allow unique definition of the toroidal spectrum of the coil current (and thus the applied RMP field).

The poloidal location, as well as the poloidal coverage of each coil, is also important, since these parameters determine the poloidal spectrum of the applied field, for a given toroidal mode number *n*. Even more importantly, with multiple rows of coils, the relative toroidal phase of the coil current between different rows - often referred to as the "coil phasing" – plays a crucial role in changing the poloidal spectrum, and therefore on the ability for the given set of coils to control ELMs. Optimization of the coil geometry, as well as the coil phasing, is one of the key aspects for both experiments and modelling. This issue will be addressed in this work. As an example, the coil currents distribution shown in Fig. 1(b) corresponds to a 90° coil phasing for the *n*=2 configuration.

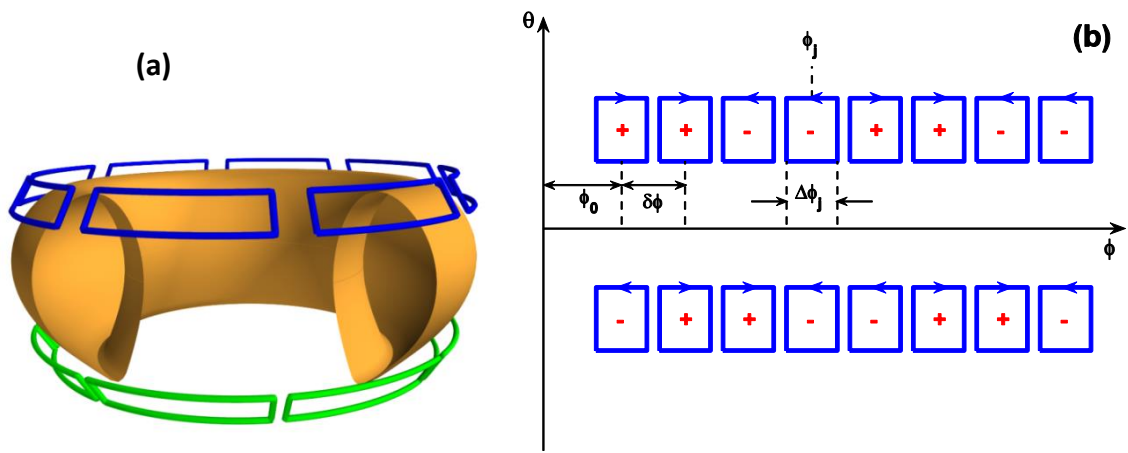

**Figure 1.** A sketch of the RMP coils distribution in (a) real geometry, and (b) on the (ϕ,θ)-plane. Shown in (b) is also an example of the coil current phasing of 90° between the upper and lower rows, for the n=2 configuration.

Since this work mainly discusses the modelling aspects of the plasma response to the RMP fields, a brief review of various response models will be given in the following section. The purpose for the plasma response modelling is two-fold. First, plasma response is essential in understanding the ELM control physics, as well as the associated transport processes. Secondly, plasma response may help us to establish certain global criteria, that can be used to guide the choice of the RMP coil current configuration in experiments, or even to guide the RMP coil design for new devices.

A key understanding of ELM control (and the accompanied density pump out) has recently been achieved, in terms of the linear plasma response to the RMP field. More specifically, a correlation has numerically been found, between the ELM controllability and the edge localized peeling response inside the plasma. This correlation seems to generally hold on various ELM control experiments in MAST [8], ASDEX Upgrade [9, 10, 11], DIII-D [12], and EAST [13], and has been shown by several codes [8, 14, 15]. This will be discussed in details in section 3, where a predictive study for ITER will also be reported based on this understanding. Section 4 summarizes the paper.

## 2. Plasma response models

There have been a few plasma response models/codes developed during recent years. Some of them were originally designed to study stellarator 3D equilibrium, e.g. VMEC [16] and HINT2 [17], some of them were developed as the extension of axi-symmetric equilibrium based MHD codes, e.g. MARS-F [18], MARS-K [19], M3D-C1 [20], and JOREK [21]. Codes such as IPEC [22] were designed specifically for studying the plasma response to 3D fields. Several of these codes have been subject to extensive joint benchmarking among themselves [23] as well as with experiments [24, 25], on the plasma response to RMP fields.

In the following, we shall briefly review some of the response models, by grouping them into four types, namely the linear response model, the quasi-linear response model, the non-linear response model, and the 3D MHD equilibrium model. Whilst the linear response model provides the so called perturbed 3D equilibrium, the non-linear MHD code can produce a dynamically accessible, often resistive, 3D equilibrium, provided that a steady state solution is achieved.

### 2.1. Linear response models

We shall discuss a range of aspects associated with the linear response models, namely single fluid versus two-fluid models, ideal versus resistive MHD models, fluid versus kinetic models, as well as the flow screening effect which plays a crucial role in understanding the plasma response to RMP fields.

2.1.1. Single fluid response

Within the single fluid framework, the response of a single resistive layer has been analytically investigated in early work [26], assuming the so called constant-$\psi$ approximation for the layer. The key understanding from this theory is that the resistive layer response is determined by three key physics factors: the plasma resistivity, the local flow speed, and the tearing index in the absence of external fields. (This tearing index is normally negative for a stable plasma.) Taking into account the average toroidal curvature effect modifies this response theory [27], though the mathematical structure of the response form is still the same.

For a toroidal plasma, two more factors also become important, that can significantly affect the linear plasma response: the geometrical effects (e.g. the RMP coil geometry, the plasma shaping etc.) and the poloidal spectrum of the applied field, for a given toroidal mode number $n$. These effects have been extensively investigated using the MARS-F model [28, 29, 30, 9, 11], which solves the linearized single fluid equations in full toroidal geometry. The MARS-F code is capable of computing either the ideal or resistive plasma response, in the presence of a generic toroidal flow (and flow shear). The RMP coil current, located in the vacuum region outside the plasma, is directly modelled in the code as the source term.

The IPEC code [22], on the other hand, computes the ideal MHD response of static plasmas. Despite the simplicity of the physics involved, the IPEC model has been shown to be hugely successful in capturing many of the macroscopic features of the plasma response as observed in experiments [31, 12]. The ideal MHD response model assumes full screening of the resonant components of the applied RMP field, even in the absence of the plasma flow. This is often not a bad approximation of reality, especially in the plasma core region, where a reasonably fast flow (say at the level of ~1% of the Alfven speed) can provide almost perfect shielding of the field [28]. Furthermore, the ideal model can also predict both edge-peeling and core-kink types of response [12], similar to that predicted by the resistive MHD model with flow [8].

2.1.2. Two fluid response

The 1D layer response model, developed by Fitzpatrick, was extended to include two fluid effects [32]. A large number of plasma response regimes were identified in this analytic model. Further analytic considerations were made, taking into account specifically the two fluid physics associated with the H-mode pedestal plasmas [33].

Two fluid physics are also included into toroidal codes such as the full MHD code M3D-C1 [20] and the reduced MHD code JOREK [21]. In terms of the plasma response, a key difference between the single and the two fluid predictions is the origin of the flow screening of the applied vacuum RMP field. According to the single fluid model, the toroidal plasma flow (essentially the thermal ion flow) has been shown to be capable of providing a significant screening of the applied field [26, 28], whilst in two fluid models, it is the electron flow perpendicular to the magnetic field line that is providing the shielding effect [34, 33]. Despite this difference in the physics origin, the mathematical structure of the screening effect, which is essentially associated with the radial component of Faraday's induction equation, is similar between the single and two fluid models [35]. It is this similarity that motivates a recent systematic study of the screening behaviour [36] within the single fluid model, when the local flow and/or flow shear approaches zero near the rational surface inside the plasma.

2.1.3. Kinetic response

Both single and two fluid models appear to produce plasma response that is in quantitative agreement with experimental measurements [37, 38, 24, 25], as long as the plasma pressure is sufficiently below the no-wall Troyon beta limit for the low-$n$ ideal kink mode. However, discrepancy occurs, in both the amplitude and toroidal phase, between the measured magnetic signals and model predictions, when the plasma pressure becomes high, i.e. close or even exceeding the no-wall beta limit [37]. This is understandable, since at high plasma pressure, the response of the pressure driven external kink mode, manifesting itself as the resistive wall mode (RWM), becomes predominant. Extensive modelling [39, 40, 41] and experimental efforts [42, 43] have led to conclusive understanding that the drift kinetic effects, due to mode-particle Landau resonances, play a crucial role in the RWM stability in high beta plasmas. Therefore, the plasma response to the external 3D fields (either RMP fields or error fields) is largely dictated by the response of the (stable) RWM under these conditions.

Indeed the MARS-K modelling results for DIII-D high beta plasmas clearly demonstrate the importance of the kinetic response in producing good agreement with experiments [44]. The agreement is obtained not only for the magnetic field perturbations measured outside the plasma, but also for the internal plasma displacement measured by soft X-rays.

Most of the ELM control experiments are carried out for plasmas well below the Troyon limit. Therefore, the fluid models are normally sufficient to produce adequate results.

*2.2. Quasi-linear response models*

Quasi-linear response models try to take into account the influence of the RMP fields on certain aspects of the plasma transport processes, e.g. the toroidal momentum transport and/or the radial particle transport. The former is important to understand RMP induced flow damping or acceleration, whilst the latter is critical to understand the so called density pump out effect observed in most of the ELM control experiments using the RMP fields. The quasi-linear models can be constructed either in a perturbative manner or self-consistently. Before discussing these two approaches, we briefly mention various types of torques that can be produced by the RMP fields, and that can affect the toroidal momentum transport.

2.2.1. Toroidal torques

Within the fluid theory, two torques can be produced due to the plasma response to the RMP fields – the Maxwell stress **jxb** torque and the torque associated with the Reynolds stress. The **jxb** torque, also often called the electromagnetic (EM) torque, is normally generated near rational surfaces for the perturbation, inside the resistive layers. Interestingly, such torque can remain finite even at the ideal MHD limit, essentially due to the continuum wave resonances. This EM torque, associated with ideal MHD, has been theoretically calculated in a slab geometry [45] and numerically verified in toroidal computations [46]. Another interesting aspect of the RMP induced EM torque is continuum resonance splitting of the radial distribution of the torque [46]. As long as the RMP field is not rotating together with the plasma (e.g. a dc RMP field), the EM torque amplitude will have several peaks near but *off* each rational surface, due to the shear Alfven wave or sound wave continuum resonances with the perturbation fields produced by RMP. The Reynolds stress torque is associated with the toroidal component of the MHD inertial term $\rho(\mathbf{v}\cdot\nabla\mathbf{v})$ in the momentum equation.

The presence of RMP fields introduces 3D perturbations to the magnetic flux surfaces, which in turn modifies the banana orbit of trapped particles. As a result, a net radial current can be produced which yields an NTV torque. The NTV torque is usually small in a tokamak plasma (with small 3D RMP field perturbations) – this is the case in the so called non-resonant NTV regime. However, the torque amplitude can be significantly enhanced by resonant effects in the particle phase space, resulting in the so called resonant NTV torque contribution. The resonance normally occurs between the toroidal precessional drift motion of trapped thermal particles and the **ExB** flow of the plasma. (The actual resonance occurs between the particle precession and the RMP perturbation as viewed in the plasma frame. Therefore, a dc RMP field should have a finite frequency of the **ExB** rotation frequency in the plasma frame.) Normally the thermal ion contribution to the resonant NTV torque dominates over the electron contribution, since the latter has much higher collisionality, which effectively removes the resonances. In fact depending on the plasma collisionality, various RMP induced NTV regimes have been identified for tokamak plasmas [47].

It should be mentioned that the precessional drift resonance is not the sole physics mechanism that produces enhanced NTV torque. The bounce resonance of trapped thermal ions, for instance, can also produce large NTV torque, under the proper (normally faster) plasma flow conditions [48, 49].

The other torque is produced by the non-ambipolar radial current associated with the "magnetic flutter" [50]. This torque has been found to largely cancel the Maxwell stress

**jxb** torque in a tokamak plasma [51]. In the presence of 3D RMP fields, in-direct torques can also be produced. For instance a net radial current can be created as a result of energetic particle losses induced by RMP. This in turn can generate a torque. So far this type of torque has not been included into the quasi-linear models for the RMP response. The magnetic stochasticity can also generate radial currents and thus produce torques as well [52, 53].

Finally, we point out that the aforementioned torques are solely due to the presence of 3D RMP fields. For the toroidal momentum balance, other (axi-symmetric) terms are also important, such as the momentum diffusion and pinch terms, and the momentum source terms due to the input injection of torques.

2.2.2. Perturbative quasi-linear model

In the perturbative quasi-linear model, the plasma response to the RMP field, at a fixed initial flow, is computed at the first step. In the next step, the perturbed quantities (magnetic field, current, velocity, etc.) associated with the plasma response are used as the input to solve the radial transport equations. Such a scheme is implemented in a recent postprocessor code RMPtran [51]. Using this code, both particle and momentum transports were simulated for a DIII-D H-mode plasma (145419), based on the M3D-C1 computed two fluid plasma response to the applied n=3 RMP field. The results reveal that, whilst the weak flow damping in that specific DIII-D discharge can be explained by the large cancellation between the accelerating non-ambipolar magnetic flutter current torque and the braking Maxwell stress torque, the density pump out effect cannot be explained based on the **ExB** type of convective particle flux utilized in the model. Additional particle flux, such as that associated with the field line stochasticity [53] and/or the NTV toque, may be important to explain this effect.

2.2.3. Non-perturbative quasi-linear model

A non-perturbative quasi-linear plasma response model has been developed and implemented into the MARS-Q code [54]. In this model, the toroidal momentum balance equation and the resistive MHD equations (i.e. the plasma response equations) are self-consistently solved as an initial value problem. This model thus can be used to simulate the RMP penetration process into the plasma, and the accompanied flow damping during the penetration.

Three torque sink terms are considered in the MARS-Q model – the Maxwell stress **jxb** torque, the Reynolds stress torque, and the NTV torque. Two models were implemented for computing the NTV torque. One is based on Shaing's semi-analytic calculations, where various collisionality regimes are smoothly connected into a unified NTV model [47]. The other is based on the MARS-K [19] computed drift kinetic energy perturbation. It turns out that the imaginary part of the drift kinetic energy, for trapped particles, is equivalent to the toroidal NTV torque [55]. These two NTV models from MARS-Q have been successfully benchmarked against each other [49], as well as against the IPEC-PENT calculations [48], for a simplified toroidal geometry. For arbitrary full toroidal NTV computations, the MARS-K approach offers more accurate results in terms of the geometrical resolution.

The self-consistent MARS-Q quasi-linear model has been applied to model the RMP induced flow damping for MAST plasmas, yielding quantitative agreement with experimentally measured flow damping [56]. The model has also recently been applied to model the RMP induced flow damping for ITER plasmas [55]. Recently, a radial transport equation is also non-perturbatively added to MARS-Q model, including the particle diffusion, pinch, ExB convective drift, as well as the particle flux due to the NTV physics. Work is in progress to use this model to interpret the RMP induced density pump out effect.

2.3. 3D equilibrium models

The 3D equilibrium models can be broadly divided into two categories – the ideal MHD model and the resistive MHD model. The ideal MHD constraint, when forced strongly, implies the existence of nested flux surfaces, but also generates singular currents. The conventional ideal 3D MHD equilibrium model is discussed in subsection 2.3.1. A possible resolution of the current singularity in the ideal MHD model has been proposed recently, that allows the existence of axi-symmetric current sheets [58]. This will be discussed in subsection 2.3.4. The resistive MHD model essentially allows the existence of field line reconnection and consequently the existence of magnetic islands. Both classical models (subsection 2.3.2) and more specialized models (subsection 2.3.3), that have recently been proposed, will be reviewed in the following.

2.3.1. Ideal MHD model

The classic ideal MHD model views a 3D equilibrium as the energy state that minimizes the sum of the plasma thermal energy and the magnetic energy

$$U = \int \left( \frac{P}{\Gamma - 1} + \frac{B^2}{2\mu_0} \right) dV$$

subject to the following constraints [59]: (i) the minimum energy state has nested magnetic flux surfaces, (ii) the (equilibrium) magnetic field is solenoidal and aligns on constant flux surfaces, (iii) the poloidal flux, enclosed by an equilibrium magnetic surface, is a function of the toroidal flux, and vice versa, (iv) the mass/entropy is conserved. Here $P$ is the plasma pressure, $B$ the modules of the magnetic field $\mathbf{B}$, $\Gamma$ the ratio of specific heats, and $V$ the plasma volume.

The above energy minimization with constraints yields a mathematically valid equilibrium solution ($\mathbf{B}$,$P$) that satisfies the force balance equation $\nabla P = (\nabla \times \mathbf{B}) \times \mathbf{B}$, as well as the condition of constant pressure along the magnetic surface.

This ideal 3D equilibrium, with nested flux surfaces, unavoidably leads to the existence of singular current density for the parallel component of the equilibrium current $\mathbf{J} = \nabla \times \mathbf{B} = \sigma \mathbf{B} + \mathbf{J}_\perp$. The neutrality condition $\nabla \cdot \mathbf{J} = 0$ immediately yields $\mathbf{B} \cdot \nabla \sigma = -\nabla \cdot \mathbf{J}_\perp$, indicating that, if the 3D equilibrium does exist, the *1/x* type of singularity will *generally* appear for the so called Pfirsch-Schlüter current, near each of the rational surfaces [60, 61]. Few exceptions are possible, and in fact may be of critical importance, as will be discussed in subsections 2.3.3 and 2.3.4.

The idea of energy minimization was numerically realized in the VMEC code [16], and its derivatives, namely the free-boundary version NEMEC [62], the extended free-boundary version ANIMEC with the possibility of treating anisotropic equilibrium pressure [63]. The energy minimization in these codes is carried out via a steepest descent method. The equilibrium solution is represented in a curve-linear flux coordinate system ($s,u,v$), with $s$ being proportional to the toroidal magnetic flux $\Phi$, $v$ the geometric toroidal angle, and $u$ a specifically chosen poloidal angle in order to optimize the poloidal spectrum bandwidth. For instance, the magnetic vector potential is generally represented as

$$A = \lambda(s,u,v)\nabla s + \Phi\nabla u - \Psi\nabla v,$$

where $\Psi$ is the poloidal flux function.

### 2.3.2. Resistive MHD model

The resistive MHD model for 3D equilibria does not assume nested flux surfaces. Instead, the MHD force balance equation is satisfied in the strong form (i.e. in differential form) by searching for the appropriate magnetic field, current and plasma pressure. To achieve this, normally an iterative procedure is required.

For instance, the PIES code [64] employs an iterative scheme to adjust the plasma current and magnetic field, starting from an initial guess for the field. Within each iteration, the pressure is calculated from the condition $\mathbf{B}\cdot\nabla P = 0$. The diamagnetic current is then computed from the force balance equation. The parallel current component is computed from the neutrality condition $\nabla\cdot\mathbf{J} = 0$. And finally, the magnetic field is re-evaluated via Ampere's law $\nabla\times\mathbf{B} = \mu_0\mathbf{J}$.

The HINT2 code [17] iteratively follows the semi-dynamic evolution of non-linear MHD equations, posed as a two-step relaxation process. At the first step, the plasma pressure is relaxed for a given magnetic field, following the condition $\mathbf{B}\cdot\nabla P = 0$, similar to PIES. At the second step, however, the simplified MHD momentum equation and Faraday's law (with plasma resistivity) is solved together, assuming a known plasma pressure. This step updates the magnetic field as well as the plasma current. The final 3D equilibrium, with the presence of magnetic islands, is obtained after achieving a converged solution of the iterative relaxation process.

2.3.3. Multi-region relaxation model

The strong singularity associated with the Pfirsch-Schlüter current, from the ideal MHD equilibrium model as discussed in subsection 2.3.1, is difficult to resolve numerically. Moreover, physically if the topology constraint is relaxed by allowing magnetic field lines to reconnect and the resulting magnetic islands to overlap, one could expect regions of radially flattened pressure profiles near rational surfaces, which eliminate the Pfirsch-Schlüter current. This leads to the idea of multi-region relaxed 3D equilibrium model [65], in which both the net helicity associated with the magnetic field lines and the mass/entropy are conserved in nested annular regions, separated by so called ideal interfaces. Minimization of the total plasma energy (thermal plus magnetic) under these two constraints leads to a partially relaxed equilibrium state, with a staircase radial profile for the equilibrium pressure. This equilibrium model thus assumes intermediate constraints between two limits – the strong ideal MHD limit conserving the field line topology, and the weak constraint limit conserving the net helicity of the whole plasma (the Taylor constraint [66]).

One of the key features of the multi-region relaxation model is the choice of the ideal interfaces which serve as the boundaries of the annular regions, in which the full relaxation takes place. These interfaces need to be sufficiently irrational (or "noble"), so that the field topology is well conserved, and in fact the ideal MHD constraint $\delta \mathbf{B} = \nabla \times (\boldsymbol{\xi} \times \mathbf{B})$ is satisfied at these surfaces. The multi-region relaxed equilibrium is numerically realized in the SPEC code [67].

Another interesting approach of the multi-region 3D equilibrium was proposed by Turnbull [68]. Different from the ideal interfaces concept in the SPEC code, Turnbull proposes the use of rational surfaces as the interfaces between nested annular regions. A helicity function is introduced along the plasma radius, which experiences jumps across the interfaces, but otherwise conserve the same helicity profile inside the annular regions, as that from the ideal MHD constraint.

2.3.4. Ideal MHD equilibria with axi-symmetric current sheets

The multi-region relaxed equilibrium model [65, 67] allows piece-wise constant equilibrium pressure that jumps across the ideal interfaces. But the model still assumes that the safety factor is continuous across the whole plasma minor radius. It has, however, recently been realized that another class of multi-region equilibria is also possible, which allows discontinuous safety factor across rational surfaces but a continuous pressure profile [58]. A discontinuous safety factor implies the existence of axi-symmetric current sheets in the equilibrium. Consequently, the equilibrium field (both poloidal and toroidal components) are also allowed to be discontinuous across the rational surfaces. One of the interesting consequence of such an equilibrium state is that the radial plasma displacement, resulted from the plasma response to the RMP field, can be continuous across the rational surface, whilst with the standard "continuous" equilibria, the radial displacement should experience a jump across the rational surface.

This new class of equilibria, found by Loizu, have several other interesting mathematical features. For instance, the Pfirsch-Schlüter current singularity, that is a problematic issue with the ideal MHD equilibrium, can now be made to disappear. Some of the unpleasant features, e.g. the discontinuous radial field perturbation [69] for such an equilibrium, can probably be eliminated by treating such an equilibrium solution as an "outer" solution, which can be matched into an "inner" layer solution where the plasma resistivity starts to play a role.

2.4. Non-linear response models

Non-linear response models rely on solving fully non-linear MHD equations (in either single or two fluid approximations) in the presence of RMP source fields. Provided that a steady state solution of the non-linear initial value problem exists, this gives a dynamically accessible 3D equilibrium.

Indeed several non-linear MHD codes have been used for computing 3D equilibria, including M3D-C1 [20], JOREK [21], and XTOR [70]. In principle, the steady state solutions of the non-linear, resistive MHD runs can recover 3D equilibria due to RMP fields, as those computed by HINT2 or PIES codes.

The 3D equilibrium, computed either by a direct equilibrium solver or as the saturated solution of the non-linear equations, will differ from the perturbed 3D equilibrium as computed by linear response codes. Therefore, it is always important to establish the validity of the linear response models. Fortunately extensive comparison with experiments demonstrates that the linear models are often adequate in describing many essential aspects of the plasma response to RMP fields [37, 8, 24, 12, 44, 11]. Besides a generic understanding that the linear models can be valid only if the applied RMP field is sufficiently small (compared to the equilibrium field), a more quantitative criterion has also be proposed, based on the idea that the linear response fails if the computed plasma displacement causes overlap of the (3D) magnetic surfaces. This condition is mathematically expressed as $|\partial \xi_n/\partial s|>1$, where $\xi_n$ is the normal displacement of the plasma, and $s$ labels the flux surface (usually the square root of the normalized poloidal flux function) [71]. As long as this criterion is not violated, the linear models provide rapid and accurate plasma response calculations to the RMP fields. In the following section, we shall show examples where the linear response MARS-F model provides key understandings to the RMP physics.

## 3. Correlation between plasma response and ELM control

In an initial work on MAST [8] of modelling the plasma response for the RMP experiments [4], it has been found that there generally exist two types of plasma response to the low-$n$ external 3D field perturbations. One is the so called core kink response, with the plasma displacement occurring globally across the whole plasma column. The other is the so called edge peeling-tearing response, where the plasma displacement is almost entirely distributed near the plasma edge. With a resistive plasma model, it is often difficult to separate the edge peeling response from the edge tearing response, though the ideal plasma model only allows the (ideal) peeling response. In later work [29,30,9,57,12], the peeling response, or the edge localized kink response, is often mentioned. We shall thus refer to the terminology of "edge peeling" response in further discussions. The key finding in Ref. [8] is a clear correlation between the fluid model predicted edge peeling response on one hand, and the experimentally observed large density pump out in both L- and H-mode plasmas in MAST and the ELM mitigation in H-mode plasmas. Since then, more modelling work has been performed on various devices [9,10,11,12,13,14,15], further confirming the basic physics that has been found in Ref. [8]. Some of the recent examples will be shown in this paper, following a discussion of the aforementioned two types of the plasma response.

*3.1. Edge peeling versus core kink response*

One example of the core kink versus the edge peeling response is shown in Fig. 2, for a MAST connected double null L-mode plasma. The MAST RMP coils consist of two sets of coils located inside the vacuum vessel, on the low field side of the torus, above and below the mid-plane, respectively. In early discharges (e.g. the one shown in Fig. 2), both sets had 6 coils equally distributed along the toroidal angle. This allows only two possible choices for the parity between the upper and the lower sets of coils – even or odd, for the n=3 configuration. Experimentally, with the same coil current of 5.6 kAt, the effect (e.g. density pump out for L-mode plasmas) of the coils on the plasma was observed only with the even parity coil configuration for discharge shown in Fig. 2.

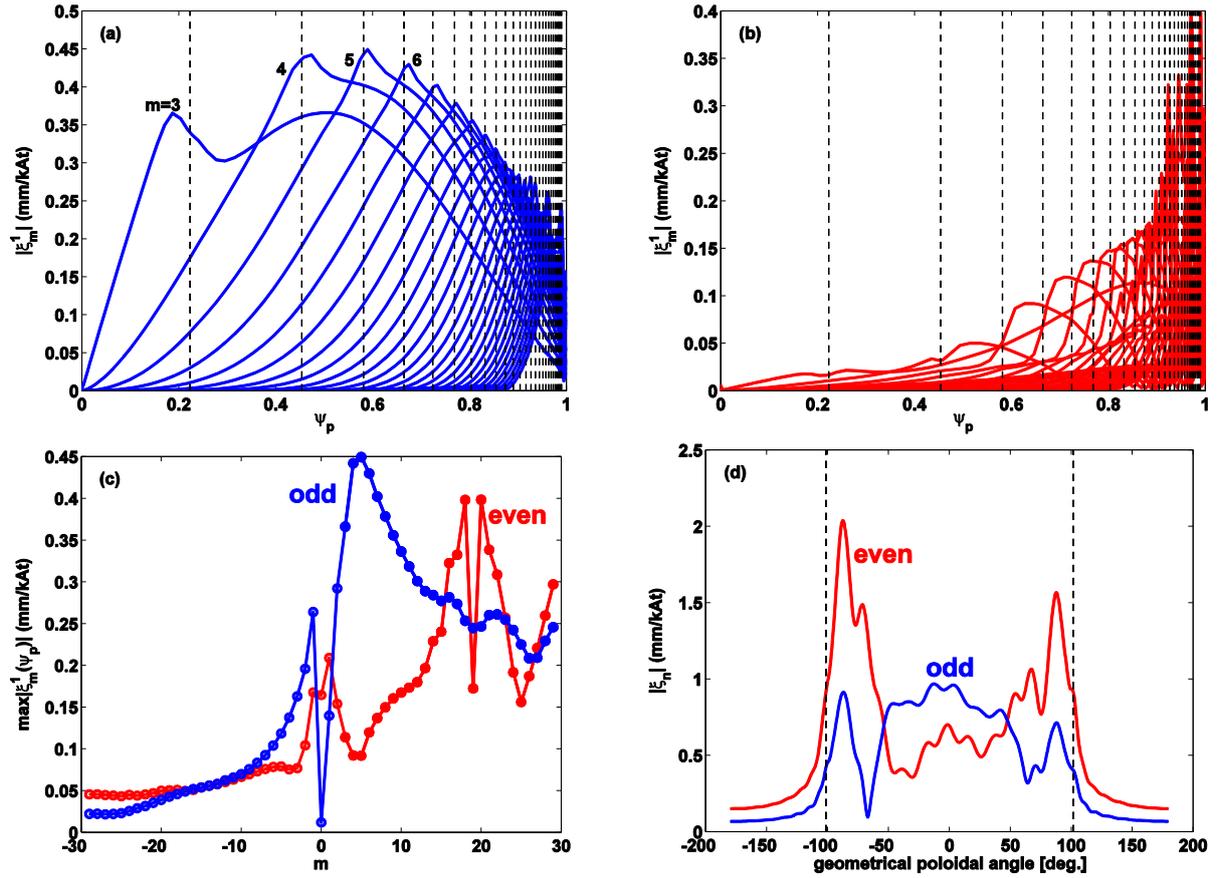

**Figure 2.** Effect of RMP coil current parity on the computed plasma radial displacement $\xi^1 \equiv \boldsymbol{\xi} \cdot \nabla s$ (with $s$ being the flux surface label): radial profiles of resonant poloidal harmonic amplitudes, triggered by (a) odd parity, and (b) even parity, coil configurations; (c) peak amplitude of poloidal harmonics over plasma minor radius; (d) poloidal distribution of the plasma surface normal displacement amplitude ($\xi_n \equiv \xi^1/|\nabla s|$). Vertical dashed lines indicate rational surfaces $q=m/(n=3)$ in (a) and (b), and indicate the X-point locations in (d). MAST connected double null, 400 kA Ohmic L-mode discharge 23001 with $q_{95}$ = 6.3 is modelled.

The MARS-F computations for this plasma show that different coil parities cause different types of plasma response, as clearly illustrated in Fig. 2 in terms of the plasma normal displacement. The odd parity coils trigger predominantly a global kink mode response from the plasma, that has large amplitude in the plasma core region. The dominant harmonics (in a straight field line coordinate system with jacobian proportional to the square of the major radius) have low poloidal mode numbers $m$ (centered around $m$=5). The plasma surface displacement, plotted along the poloidal angle, tends to peak near the outboard mid-plane, exhibiting the low-$n$ kink-ballooning characteristics. On the contrary, the even parity coils trigger predominantly an edge peeling response, having large amplitude near the plasma edge region. The dominant poloidal harmonics are centered around $m=20 (\sim nq_{95})$. This type of plasma response has a different poloidal distribution at the plasma boundary, namely the surface displacement now tends to peak near the top and bottom of the torus, i.e. near the equilibrium X-points.

This computational finding of the X-point displacement prompted further experimental study of the plasma behaviour near the X-point in MAST plasmas, leading to the clear observation of lobe structures [72], which are believed to be formed by stable and unstable separatrix manifolds near the X-point [73]. It remains, however, to be understood what the intrinsic relation (if any) is between these lobe structures and the X-point displacement as predicted by the fluid theory.

Below we provide an heuristic understanding of why the core kink (edge peeling) response results in outboard mid-plane (X-point) displacement of the plasma boundary. The core kink response usually consists of a large number of poloidal harmonics, having similar amplitude near the plasma edge (see e.g. Fig. 2(a,c)). This often yields a ballooning type of plasma displacement, peaking at the low field side of the torus. The edge peeling response, on the contrary, involves a smaller number of poloidal harmonics with large amplitude. More importantly, some (often one) of the harmonics closely satisfy the resonance condition $q=m/n$. On the other hand, the following relation holds in a generic toroidal geometry (for an ideal plasma), between the perturbed radial magnetic field (flux) $b^1$ and the normal displacement $\xi_n$

$$b^1 \equiv \frac{q}{\mathbf{B}\cdot\nabla\phi}(\delta\mathbf{B}\cdot\nabla\psi) = \left[\frac{\partial}{\partial\chi} - q\frac{\partial}{\partial\phi}\right](RB_p\xi_n),$$

where $B_p$ is the equilibrium poloidal field. If a single, near-resonant harmonic becomes dominant in the plasma response, the above relation yields $\xi_n \sim b^1/B_p$, meaning that the normal displacement will become large at the poloidal angle, where the poloidal equilibrium field is weak, which is the case for the X-point. But a weak poloidal equilibrium field does not have to be associated with an X-point. An example is shown in Fig. 3, where we show the MARS-F computed normal displacement of the plasma (Fig. 3(b)) for a lower single null plasma from ASDEX Upgrade. The displacement in fact peaks near the top of the torus as well in this case. This is despite the rather different poloidal distribution of the perturbed radial field $b^1$, shown in Fig. 3(a).

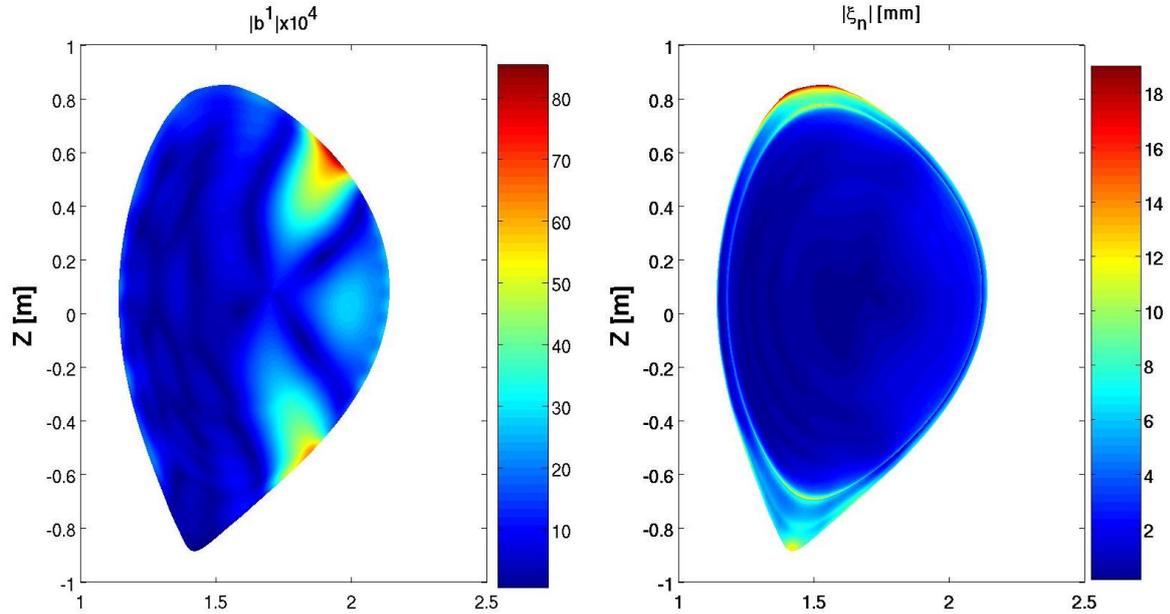

**Figure 3.** MARS-F computed perturbed radial field amplitude including the plasma response (left panel, normalized by a factor $R_0^2 B_0$) and plasma normal displacement amplitude (right panel), for an equilibrium constructed from ASDEX Upgrade shot 33133 at 2600ms.

The RMP coil configuration is not the only factor affecting the plasma response mode as well as the density pump out and/or the ELM behaviour in experiments. Generally speaking, any factors that change the spectrum of the applied field can play a role, e.g. the coil geometry, the plasma shaping, the safety factor $q$, etc. It appears that the safety factor near the plasma edge ($q_{95}$) is particularly important, by affecting the resonant portion of the Fourier spectrum. This has been demonstrated in experiments [4, 74]. The MARS-F computations also show that, with the same coil current configurations, either core kink or edge peeling response can be triggered, depending on $q_{95}$. One example from MAST is shown in Fig. 4. At the higher $q_{95}$, the RMP fields trigger a core kink mode type of plasma response. At the lower $q_{95}$, the plasma response tends to peak near the edge, with dominant poloidal components at higher $m$ numbers. The plasma surface displacement peaks at the outboard mid-plane at $q_{95}$ = 5.34, where no density pump out is observed in the experiments. At $q_{95}$ = 4.51, where a strong pump out effect is observed, the computed surface displacement peaks near the X-points. In fact, by scanning $q_{95}$ from 5.45 down to 4.51, we find a smooth transition from the mid-plane peaking to the X-point peaking.

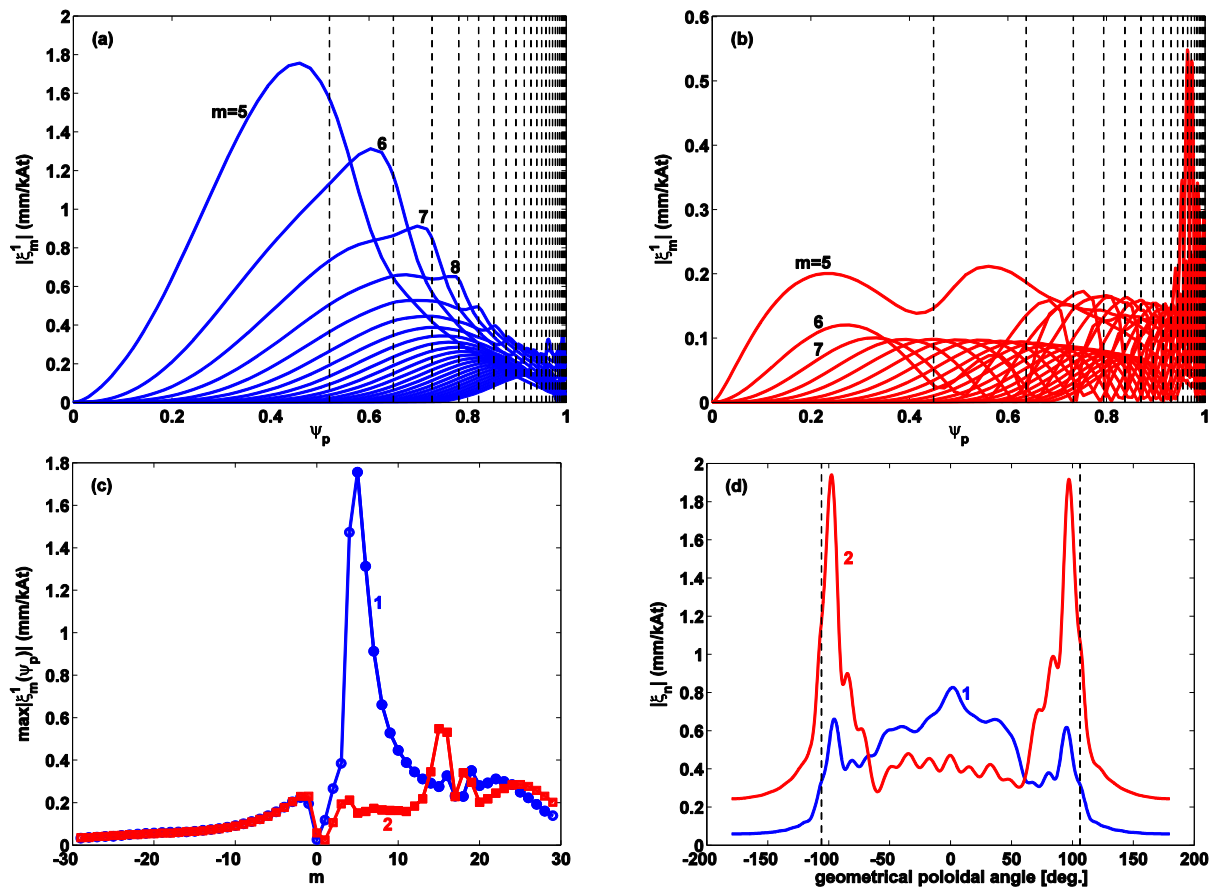

**Figure 4.** Effect of $q_{95}$ on the computed plasma normal displacement: radial profiles of resonant poloidal harmonic amplitudes with (a) $q_{95}$ = 5.34 and (b) $q_{95}$ = 4.51; (c) peak amplitude of poloidal harmonics over plasma minor radius; (d) poloidal distribution of the plasma surface displacement amplitude. The curve labelled '1' ('2') in figures (c) and (d) corresponds to $q_{95}$ = 5.34 (4.51). Equilibria are based on MAST H-mode discharge 24460. Odd parity coils are assumed. Density pump out is (not) observed in experiments as $q_{95}$ below (above) 5.

With the same RMP spectrum, other plasma conditions may also affect the edge peeling versus the core kink response, as demonstrated by Fig. 5, where the difference in the toroidal flow speed near the plasma edge plays a decisive role. The response of a 900 kA L-mode MAST plasma to the $n=3$ odd parity coil currents is computed. Unlike the case shown in Fig. 2, odd parity coils cause the density pump out in this discharge, but not the even parity. MARS-F computations generally show the appearance of the peeling-like response with the odd parity coil configuration for this case. However, the peeling components are much more pronounced using the rotation profile measured during the RMP currents on, than that using the rotation profile measured from a similar discharge, but with the RMP currents off. These two rotation profiles differ mostly near the plasma edge region. These results suggest a possible scenario of density pump out, that the RMP fields give an initial braking of the plasma edge rotation, leading to further RMP field penetration and the triggering of the peeling-like plasma response, that eventually causes the density pump out. The experimental observation, that the density pump out occurs only at sufficiently large coil current, is possibly related to the braking of the edge rotation.

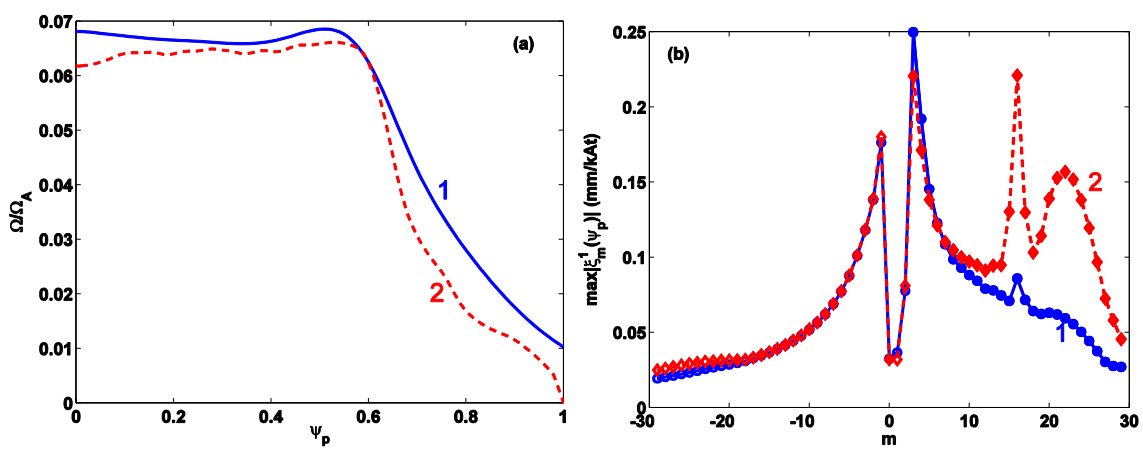

**Figure 5**. Effect of toroidal rotation profile on the triggering of the peeling mode response: (a) measured rotation profiles without the RMP coil currents (labelled '1') and during the application of RMP (labelled '2'); (b) peak amplitude of poloidal harmonics over plasma minor radius. The numbering of the curves in (b) corresponds to that in (a). MAST L-mode discharge 24534 with $q_{95}$ = 5.9 is modelled, with odd parity RMP coils.

## 3.2. Modeling versus experiments

Since the initial finding of the correlation between the MARS-F computed edge peeling response and the experimentally observed density pump out in MAST, for both L- and H-mode plasmas [8], further evidence has been found from other devices during recent years. Whilst for MAST plasmas, the best figure of merit appears to be the ratio of the X-point displacement to the outboard mid-plane displacement, as shown in Fig. 6, we find that for other devices, the X-point displacement itself is often the best measure. Moreover, the magnitude of the pitch resonant radial field perturbation near the plasma edge has also been found to be a good indicator for the observed RMP effects on ELMs [74, 10], provided that the radial field perturbation includes the plasma response.

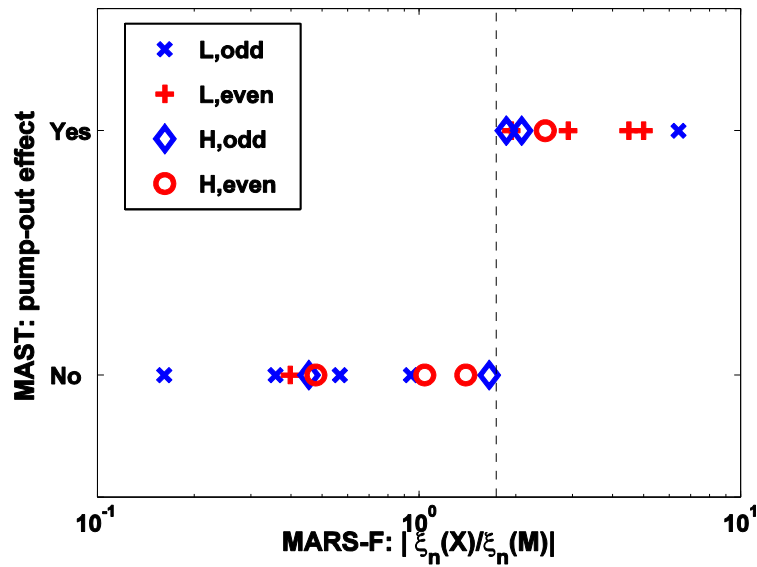

**Figure 6.** Correlation between the computed normal displacement of the plasma surface, and the observed density pump-out effect in all types of MAST plasmas from the RMP experiments. Plotted is the experimental density pump-out effect versus the ratio of the displacement peaking near the X-point ($\xi_n(X)$) to that at the outboard mid-plane ($\xi_n(M)$).

One example from ASDEX Upgrade is shown in Fig. 7, where an $n=2$ RMP field, with time-varying coil phasing, is applied to a high $q_{95}$ plasma ($q_{95}=5.5$). Clear ELM mitigation (a), as well as the density pump out (b), is observed for coil phasing (defined as the toroidal phase difference between the upper and lower rows of the $n=2$ coil current) at about -100° (c-d). On the other hand, the MARS-F computed plasma response, measured either in terms of the last resonant radial field amplitude near the plasma edge (solid line in (e)), or in terms of the X-point displacement (solid line in (f)), reaches its maximum at this coil phasing. Neither the vacuum field criterion (dashed line in (e)) nor the outboard mid-plane plasma surface displacement (dashed line in (f)) provides the proper indicator for the experiments.

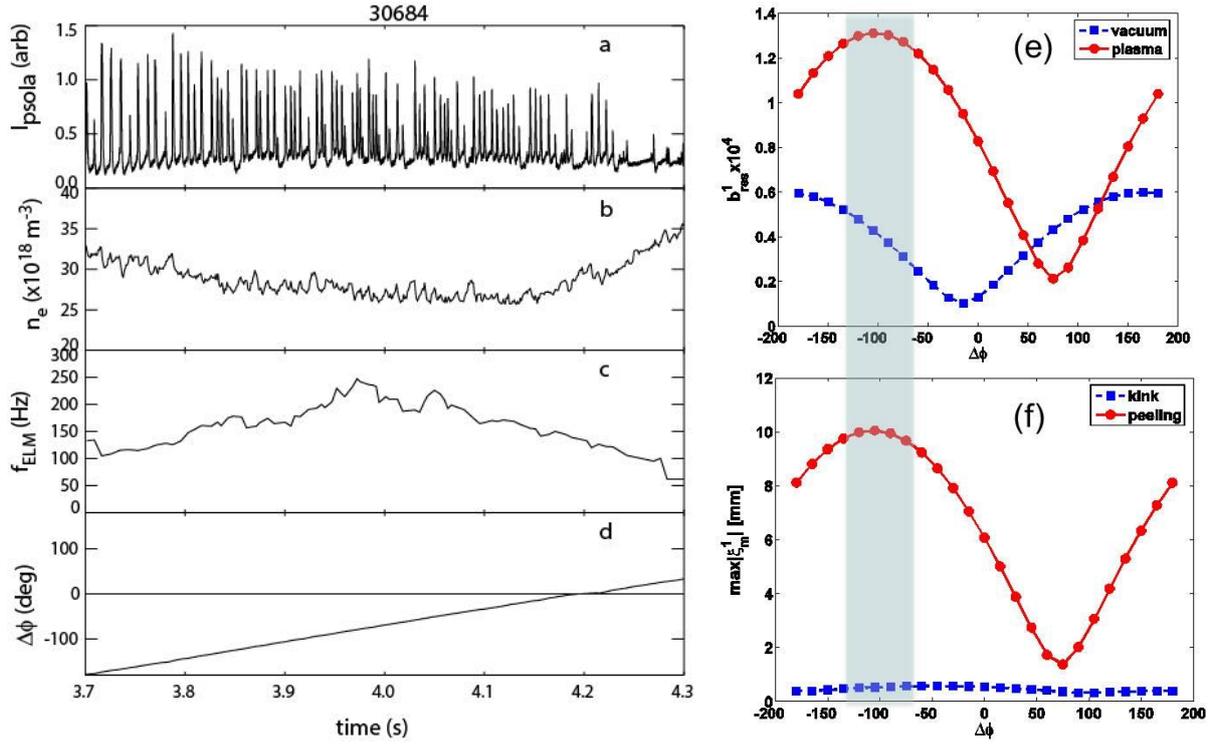

**Figure 7.** The ASDEX Upgrade RMP discharge 30684 (a-d) versus the MARS-F modelling results (e-f), for the coil phasing scan with the n=2 configuration: (a) the type-I ELM behaviour as detected by the divertor saturation current, (b) the line averaged electron density, (c) the ELM frequency, and (d) the coil phasing $\Delta\Phi$ between the upper and lower rows of RMP coils, (e) the computed last resonant field amplitude (vacuum versus total plasma response), and (f) the magnitude of the computed kink versus peeling components. The shaded regions correspond to the optimal coil phasing.

Recent modelling results [13] for the ELM control experiments in EAST [7] also confirm the correlation between the edge peeling response and the observed ELM mitigation/suppression. We show one example in Fig. 8. EAST also has two rows (above and below the low field side mid-plane) of RMP coils installed, each consists of 8 coils. In discharge 56360, the *n*=2 coil configuration was used, with two values of coil phasing, referred to as 270° and 90°, respectively (Fig. 8(a)). During the 270° phase (3.1-3.8 s), the frequency of the type-I ELMs is significantly increased (b), as compared to the pre-RMP phase. The ELM mitigation phase is accompanied by large density pump out and strong toroidal flow damping (c). On the contrary, the 90° phase (3.8-4.6 s) leads to only a slight increase of the ELM frequency, almost no effect on the plasma density and stored energy, and a minor damping on the plasma flow.

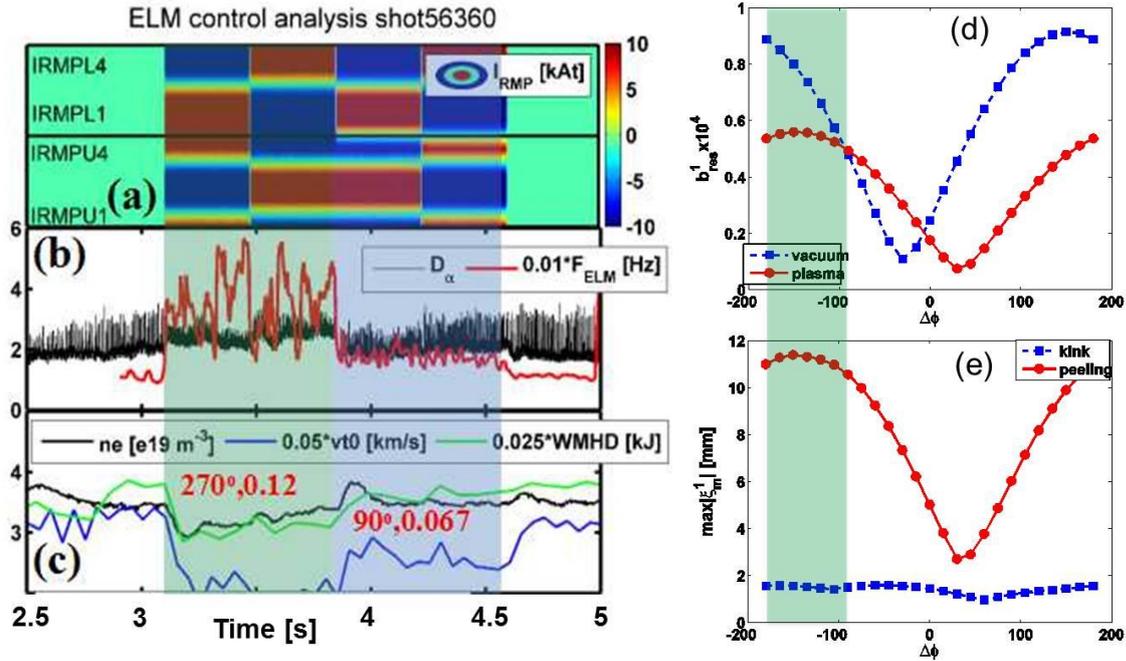

**Figure 8.** The EAST RMP discharge 56360 (a-c) versus the MARS-F modelling results (d-e), for the coil phasing scan with the n=2 configuration: (a) the lower and upper rows of coil current setup, (b) the type-I ELM behaviour as detected by the $D_\alpha$ signal, as well as the estimated ELM frequency, (c) the line averaged electron density, core toroidal rotation speed, and the thermal energy content, (d) the computed last resonant field amplitude (vacuum versus total plasma response), and (f) the magnitude of the computed kink versus peeling components. The shaded regions in (a-c) (between 3.1-3.8s) and (d-e) correspond to the optimal coil phasing.

The MARS-F modelling results are shown in Fig. 8(d-e). The amplitude of the last resonant radial field component including the plasma response (solid line in (d)), as well as the amplitude of the peeling response (solid line in (e)), both show larger response at the coil phasing of $\Delta\phi=-90^o$ (i.e. $270^o$) than that of $+90^o$. The vacuum field (dashed line in (d)), on the other hand, would predict a larger effect at $+90^o$ coil phasing. The core kink response (dashed line in (e)) is generally weak for this EAST plasma. The MARS-F results also predict that even better ELM mitigation can be achieved with the coil phasing of about $-150^o$ (i.e. $210^o$). This can be experimentally validated in the future, with the installation of more power supplies to the RMP coils in EAST.

### 3.3. Prediction for ITER

Based on the edge peeling response criterion, modelling work has been carried out for ITER plasmas [75], in particular for the 15 MA Q=10 inductive scenario [76], with one example shown in Fig. 9. Here we assume the RMP coil system as designed for ITER – consisting of 3 rows of coils along the poloidal angle at the low field side, each having 9 coils uniformly distributed along the toroidal angle [77]. We also assume a 45 kAt coil current amplitude in each row (at a given toroidal mode number $n$). This current is half way below the ITER design limit of 90 kAt. On the other hand, the results shown can be easily scaled to the full current, since these are from the linear response computations. Although most likely, higher-$n$ coil configurations will be considered for the ELM control in ITER (to avoid mode locking with low-$n$ fields), we compute the plasma response to $n$=1-4 RMP fields, and show the results in Fig. 9(a-d), respectively. For each $n$, we vary the relative toroidal phase of the coil currents in the upper and lower rows, with respect to the middle row currents. We then compute the X-point displacement of the plasma scanning the 2D parameter space. The results show the existence of one single optimal coil phasing, for each $n$, that maximizes the X-point displacement, or equivalently maximizing the edge peeling response. Interestingly, the optima are not far from each other for the $n$=3 and $n$=4 configurations.

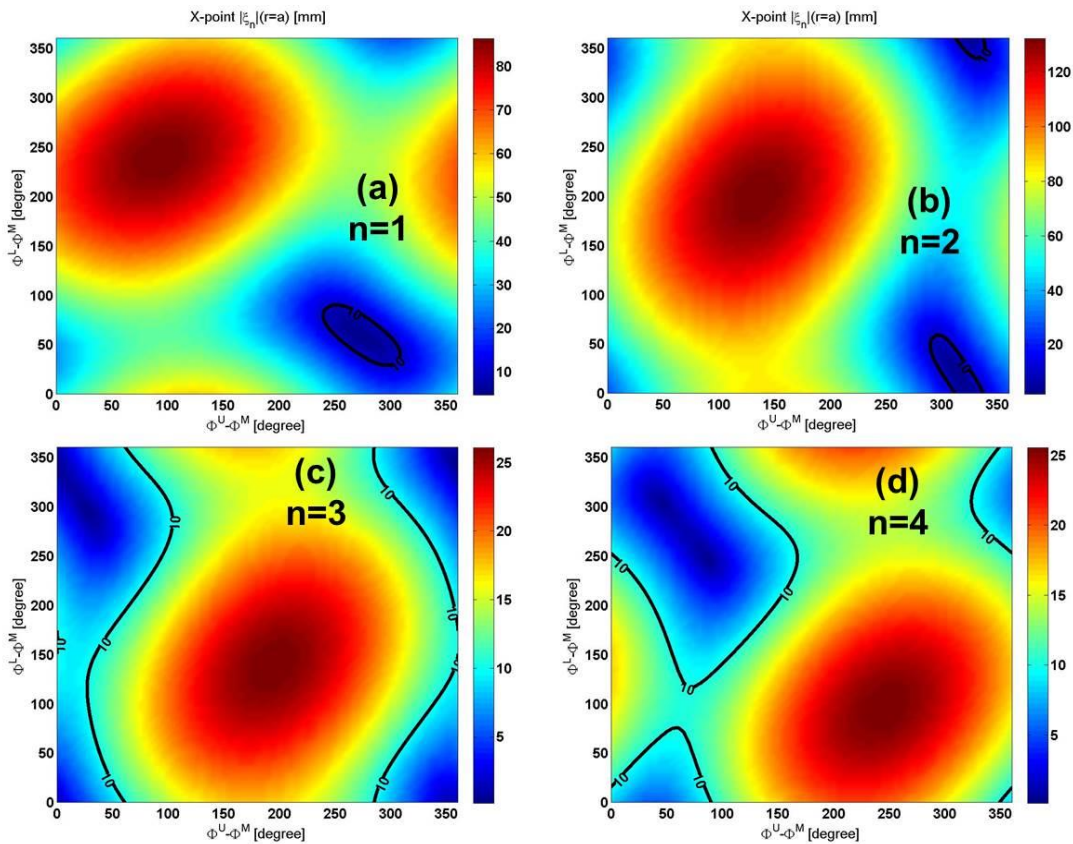

**Figure 9.** The MARS-F computed X-point displacement for the ITER 15MA Q=10 inductive scenario plasma, assuming $n$=1,2,3,4 RMP coil configurations, respectively. Three rows of the ITER ELM control coils are used, with varying coil phasing between the upper and middle rows (horizontal axis), and between the lower and the middle rows (vertical axis). The coil current amplitude is assumed to be the same (45 kAt) for all three rows, for each given $n$.

The plasma response based criteria can also be utilized to guide the RMP coil design for future devices. Such an exercise has recently been carried out [78], again on an example of the ITER plasma (despite the fact that the coil design has been fixed for ITER). Figure 10 shows sample results, where we compute the $n$=4 plasma response, measured by the amplitude of the last resonant radial field component, while either scanning the poloidal location $\theta_c$ of the center of the upper and lower rows (symmetrically) of coils (at fixed coil width as from the ITER design, left panel), or scanning the coil width $\Delta\theta$ of both rows (at fixed coil location as from the ITER design, right panel). The coil width is defined here as the coverage of geometric poloidal angle by each coil. The vacuum field (dashed lines) is compared with the total field perturbation including the plasma response (solid lines). Only the upper and lower rows are considered in this example, with three choices of the coil phasing $\Delta\Phi$=0°, 90°, 180° shown in Fig. 10. The vertical lines indicate the corresponding ITER design values (which are slightly different between the upper and lower rows). Figure 10 shows that, in terms of maximizing the total response field amplitude, the coil location of the ITER design is close to the optimum for the $n$=4 field in even (or close to even) parity, but is not optimal for the $n$=4 field in odd parity configuration. The ITER designed coil width for the off-midplane rows, on the other hand, is sub-optimal for all choices of the coil phasing. The coils poloidal width in ITER has been limited in the design due to the need to also incorporate the in-vessel vertical stability coils [77]. On the other hand, additional study shows that the size of the ITER middle row coils is very close to the optimum for the $n$=3 and $n$=4 configurations [71]. Therefore, combinations of all three rows of RMP coils, with the present ITER design, should be capable of offering sufficient flexibility and providing optimal fields for the ELM control.

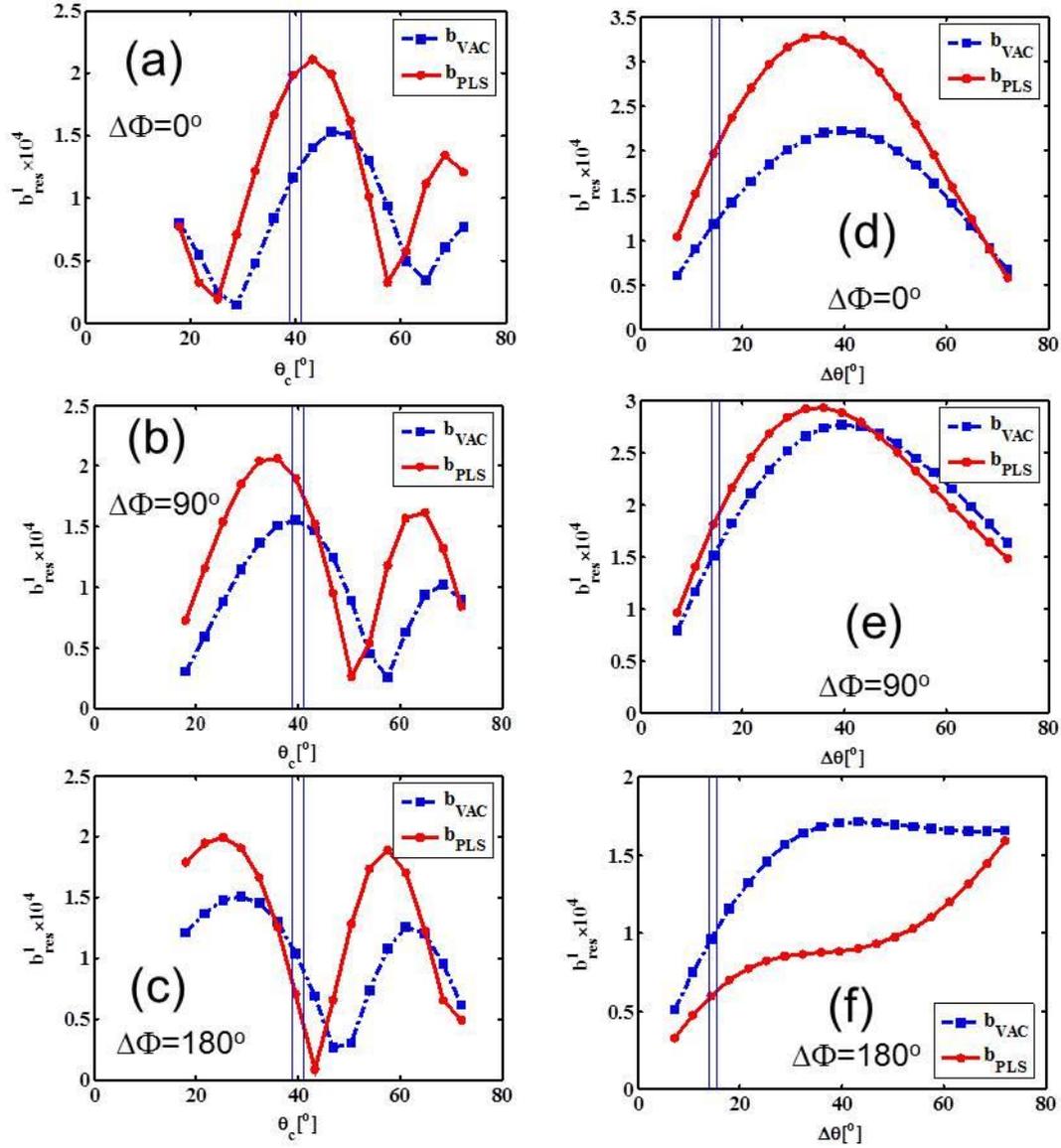

**Figure 10.** Comparison of the amplitude of the last pitch resonant radial field component, in [Tesla], between the vacuum field (dashed) and the total field including the plasma response (solid), versus the coils' poloidal location $|\theta_c|$ (left panel, at fixed $\Delta\theta=15.5°,14.1°$) and the coils' width $\Delta\theta$ (right panel, at fixed $\theta_c=39.0°,-41.1°$, for the toroidal phasing of the coil currents at (a,b) $\Delta\Phi=0°$ (even parity), (c,d) $\Delta\Phi=90°$, and (e,f) $\Delta\Phi=180°$ (odd parity), respectively, using upper and lower rows of coils in the *n*=4 configuration with 45 kAt current. Vertical lines indicate the ITER design values for the coils ($\theta_c^U=39.0°$, $-\theta_c^L=41.1°$, $\Delta\theta^U=15.5°$, and $\Delta\theta^L=14.1°$). Considered is a reference plasma from the ITER 15 MA Q=10 inductive scenario.

## 4. Summary and discussion

This paper briefly reviews various plasma response models that have presently been considered in the context of the ELM control using the 3D RMP fields produced by magnetic coils external to the plasma. These models can crudely be categorized into 4 groups (with representative computational tools/codes): (i) the linear response models (MARS-F/K, IPEC, linear version of M3D-C1), resulting in perturbed 3D equilibrium; (ii) the quasi-linear response models (MARS-Q, RMPtran), that couple the plasma response to certain aspects of the plasma transport processes associated with RMPs, e.g. the toroidal momentum and the plasma particle transport; (iii) the full 3D equilibrium models (VMEC, HINT2, PIES, SPEC), that require the satisfaction of MHD force balance equation, either in the strong form or in the weak form, assuming either ideal or resistive plasma description; (iv) the full non-linear MHD models (M3D-C1, JOREK), that can in principle lead to dynamically accessible equilibria, often with the presence of magnetic islands.

Naturally, more advanced models normally involve more physics and often much more challenging computing. Simpler models, e.g. the linear response models, can sometimes be advantageous in offering fast computing, systematic investigations in multiple parameter space, easier access to understanding the physics. But most important of all, linear response models are experimentally validated to yield reasonable results that agree with measurements. The key reasons for this validity are (i) the relatively slow time scale (often much slower than the MHD time scale) involved in the RMP problems; (ii) the large spatial scale of the applied field structure, since often low-$n$ fields are applied for ELM control; and (iii) the amplitude of the applied field being several orders of magnitude smaller compared to the plasma equilibrium field. The $3^{rd}$ condition is necessary but not sufficient, since even a small perturbation can give non-linear effects, if the resulting plasma displacement causes overlapping of flux surfaces. Finally, the linear, macroscopic plasma response fields can be used as input to other codes such as XGC0 [79] for more advanced (kinetic) physics modelling.

Given the fact that the physics of the RMP induced ELM mitigation/suppression is still far from fully understood, it is always desirable for theory to offer certain simple criteria that can be used to guide the experiments. The initial criteria (e.g. the so called vacuum-island-overlap-window [80]), based on the vacuum field approximation, have shown a correlation with ELM suppression on some devices, but does not provide a sufficient condition to ensure suppression. A new criterion, based on the edge peeling response, appears to be capable of explaining experimental results in several devices, as have been shown by a series of examples included in this paper, as well as by recent studies using the full 3D equilibrium model [15] or the non-linear MHD model [14]. This criterion thus may offer an interesting alternative guide for the RMP coil design and optimization for future devices. On the other hand, the physics understanding of the correlation between the fluid model predicted peeling response on one side, and the experimentally observed ELM mitigation/suppression and/or density pump out on the other side, is still not complete. The peeling response criterion itself may still need refinement, in particular towards more quantitative prediction (e.g. of the amplitude of the applied RMP fields). Nevertheless, this criterion seems to be a promising alternative to the vacuum field based criteria. Finally, we remark that it appears that more subtle physics are involved in distinguishing between the ELM mitigation and suppression. This is still an active research area.

## Acknowledgements

This work has been carried out within the framework of the EUROfusion Consortium and has received funding from the Euratom research and training programme 2014–2018 under grant agreement No 633053 and from the RCUK Energy Programme [grant number EP/I501045]. Work is also part funded by National Natural Science Foundation of China (NSFC) [grant number 11428512] and by National Magnetic Confinement Fusion Science Program under grant No. 2015GB104004. The views and opinions expressed herein do not necessarily reflect those of the European Commission. ITER is the Nuclear Facility INB no. 174. The views and opinions expressed herein do not necessarily reflect those of the ITER Organization.